# Automatic Detection of Public Development Projects in Large Open Source Ecosystems: An Exploratory Study on GitHub

Can Cheng, Bing Li, Zengyang Li*, Peng Liang
School of Computer Science
Wuhan University
Wuhan, China

*Abstract* — Hosting over 10 million of software projects, GitHub is one of the most important data sources to study behavior of developers and software projects. However, with the increase of the size of open source datasets, the potential threats to mining these datasets have also grown. As the dataset grows, it becomes gradually unrealistic for human to confirm quality of all samples. Some studies have investigated this problem and provided solutions to avoid threats in sample selection, but some of these solutions (e.g., finding development projects) require human intervention. When the amount of data to be processed increases, these semi-automatic solutions become less useful since the effort in need for human intervention is far beyond affordable. To solve this problem, we investigated the GHTorrent dataset and proposed a method to detect public development projects. The results show that our method can effectively improve the sample selection process in two ways: (1) We provide a simple model to automatically select samples (with 0.827 precision and 0.947 recall); (2) We also offer a complex model to help researchers carefully screen samples (with 63.2% less effort than manually confirming all samples, and can achieve 0.926 precision and 0.959 recall).

## I. INTRODUCTION

In recent years, the GitHub ecosystem has witnessed an increasing popularity, and it has attracted more than one hundred studies focused on it [1]. With more than 10 million of software projects hosted on this ecosystem, it is hard to select appropriate samples (i.e., projects) when conducting large scale case studies. Early studies often use manual selection methods to select samples. But as large datasets appear [2], researchers often face a dilemma that they want to use a large dataset to verify the generality of their results and at the same time they cannot confirm whether these projects meet their research goals. Thus, it is important to find a way to automate the sample selection process.

Most research on GitHub needs to satisfy an implicit hypothesis: their sampled projects under investigation are public development projects, which means that these projects should be open to public and the content of these projects should be about development. For example, when studying communities and teams of projects in GitHub, their samples must be public development projects. Because many projects hosted on GitHub are not for software development (e.g., blogs, translation and student homework) and are private [3]. When the sample size is small, it is doable to read the descriptions and readme files manually to select appropriate samples. However, as the dataset becomes larger, this manual method becomes inefficient. Hence, in this study, our goal is to automatically detect public development projects.

We developed a model in this work to automatically detect public development projects based on the J48 decision tree algorithm [4]. We verified our model on a dataset of 6,715 GitHub projects labeled by master and PhD students on software engineering. Our model performs well in classifying public development projects. The main contributions of this work are:

- We identified a set of words and phrases (e.g., mirror, personal) that can reflect projects' properties.
- We fitted a simple decision tree that can classify public development projects with a precision of 0.839 and a recall of 0.950, and this model can help researchers effectively select appropriate samples.
- For those studies that have strict requirements on the dataset, we provide a complex decision tree, with 63.2% less effort than manually confirming all samples, and it can obtain a classification with a precision of 0.926 and a recall of 0.959.

In the rest of this paper, related work is discussed in Section II. Design of this study is described in Section III. The results are elaborated in Section IV. Threats to validity of the results are presented in Section V, and this work is concluded in Section VI.

## II. RELATED WORK

### A. Problems in Studying Open Source Ecosystem

Nowadays, more and more research studied software ecosystems [5] (most objects of software ecosystem research are open source software (OSS) ecosystems), the potential reasons for this phenomenon is that open source ecosystems provide publicly available historical datasets which researchers can benefit from.

However, there are some problems when studying the historical data of open source ecosystems. Howison and Crowston found that projects hosted on SourceForge were often abandoned and their information was often missing since some project data are hosted outside SourceForge [6]. Weiss argued that it is not necessary to consider all SourceForge data because of the fickleness of some projects [7]. Rainer and Gale conducted in-depth analysis on the quality of SourceForge data [8]. They noted that only 1% of SourceForge projects were actually active, and suggested that researchers should be careful when using project samples. Kalliamvakou *et al.* investigated on the defects of GitHub datasets [3, 9]. They detected 13 perils and gave strategies to avoid these perils.

Although researchers have already been aware of the problem, some solutions to the problem still remain in the stage of manual verification. When facing a rapidly-increasing amount of data on OSS projects, these methods are not effective any more.

*B. Studies on GitHub Datasets*

As our dataset was retrieved from the GitHub ecosystem [10], we first investigate the datasets which can be potentially used to study GitHub. Cosentino *et al.* conducted a systematic mapping study on GitHub [1] and concluded that currently there are six ways to get GitHub data: (1) GHTorrent [2], (2) GitHub Archive, (3) GitHub API, (4) others (e.g., BOA [11]), (5) manual approach, and (6) a mixture of them. It is pointed out by Kalliamvakou *et al.* [3] that GitHub Archive started data collection in 2011, and is an incomplete mirror to GitHub. GHTorrent, in comparison, has retrieved the complete history of GitHub. Moreover, GHTorrent can be extended by GitHub API which is most frequently used, hence we collected studies based on GHTorrent.

We selected high quality articles (published in top journals/conferences or highly cited) that used GHTorrent to see how researchers chose samples to avoid potential threats. The results are shown in TABLE I.

TABLE I. SAMPLE SELECTION METHODS IN LITERATURE

| Method | Literature |
|---|---|
| Remove projects that have poor perfomance in some dimensions (e.g., pull request) | [12-18] |
| Select projects that are top in some dimensions (e.g., star number) | [19-21] |
| Only consider projects in several programming languages (e.g., ruby) | [13, 14, 18] |
| Remove projects that are forks | [17, 18] |
| Use a task-related strategy (e.g., select .rb files in ruby projects) | [12, 13, 17, 18, 21] |
| Have no clear project selection strategy | [22-25] |

As shown in TABLE I, sample selection methods are not unified in literature. Most of the literature aims to select samples according to specific criteria (e.g., projects that are top in number of stars), which reduces the diversity of samples. In addition, these sample selection methods also do not distinguish between development projects and the projects for other purposes, like storage. Intuitively, development projects and projects for other purposes are developed in different ways, and this will negatively affect the validity of research results. In this work, we tried to solve this problem.

### III. STUDY DESIGN

As our goal is to automatically detect public development projects, and standard datasets are required to validate our proposed model. In this section, we discuss how to create a standard dataset and how to automatically detect public development projects.

*A. Key Concepts*

In this study, the key concept is public development project, thus we first give its definition in two aspects:

**Public project**: Projects that are not built for private use. Anyone has the opportunity to participate in these projects.
**Development project:** Projects that are set up for software development. This type of tasks include repositories of libraries, plugins, gems, frameworks, add-ons, and so on [3].

*B. Research Questions*

We formulate the following research questions (RQs) to investigate the feasibility of automatic detection of public development projects.

**RQ1**: Can public development projects be detected automatically without collecting additional data?

**Rationale**: There are many datasets (e.g., GHTorrent) and tools (e.g., GitHub API and web crawlers) that can be used to collect OSS data. However, although some methods (e.g., web crawlers) can get rich information, these methods are inefficient due to access restrictions of GitHub to fixed IP addresses. In order to increase the usability of our approach, we decided to do this work with only readily available data. In this study, we only use data from GHTorrent.

**RQ2**: How well can public development projects be detected?

**Rationale**: Existing sample selection methods have either a good recall rate (e.g., removing poor projects) or a good precision rate (e.g., selecting top projects). These methods have their own flaws: (1) inaccurate samples lead to inaccurate conclusions; (2) missing samples result in poor generalizability of the conclusions. In this study, we aim to automatically select public development projects with both high precision and recall rates.

*C. Standard Dataset*

Creating a standard dataset means that we need to know which projects are public development projects. Since we cannot obtain a publicly available dataset that contains such information, we decided to manually create such a dataset (i.e., extending the GHTorrent dataset).

In order to make the dataset more convincing, the size of our dataset should not be too small. To achieve this goal we asked several master and PhD students on software engineering to manually examine which projects are public development projects. Due to the limited resource, we could not mobilize many people to confirm all projects hosted on GitHub, which contains more than 10 million projects. Therefore, with limited samples that we can manually check, we should create a strategy to ensure that our samples contain as many types of projects as possible.

As the number of projects that we can manually check is limited (around 10,000 projects according to our available resource), we should select 10,000 samples from over 10 million projects. Then there are two possible strategies: (1) Select all projects created over a period of time; (2) Randomly select a specific number of projects. Considering that a random selection is unstable (i.e., different selection results may contain different types of projects), we decided to select all projects created over a period of time in this study.

After selecting samples, we need participants to decide whether a sample is a public development project. This process may introduce personal bias. In order to alleviate this problem, we first defined how to identify public software development projects (see TABLE II). Then, we randomly selected 100 samples and strictly labeled these samples according to identification criteria in TABLE II. We showed these labeled samples to participants to give them a first impression to reduce their personal bias.

TABLE II. IDENTIFICATION CRITERIA OF PUBLIC PROJECTS AND SOFTWARE DEVELOPMENT PROJECTS

| Category | Identification |
|---|---|
| Public project | If the project's description and readme file do not state that the project is a **private project** (i.e., this project is established for the project owner's own use), we classify this project as a public project. |
| Software development project | We classify projects as **software development projects** if their contents are files used to build tools of any sort. This type of use includes repositories of libraries, plugins, gems, frameworks, add-ons, and so on [3]. |

Lastly, we assign tasks to participants according to the following procedure (see Fig. 1):

1) Divide the set of samples into several subsets.
2) For each sample subset, assign a participant to decide whether each sample meets the criteria in TABLE II. If participants are not sure which category a sample belongs to, they can put such a sample in a collection "undecided" first.
3) Collect all undecided samples, and convene all participants to discuss how to categorize these samples. For each sample, participants need to give their classification and reasons. If a consensus cannot be achieved, we vote to decide which category the project should be assigned to .
4) Form the final dataset by combining clearly categorized samples based on the discussion.

In conclusion, we first collected samples created over a period of time in GitHub. Then we assigned these samples to participants for manual classification. Finally, by summarizing these results, we got the final standard dataset.

*D. Automatic Detection of Public Development Projects*

After getting the standard dataset, the next step is to automatically classify these samples. In the GHTorrent dataset, the project-related data include issue data, commit data, committer data, and basic data of the project (owner, description, star number, watcher number, language, and readme file). Hence, we collected 1,000 samples (in the sample set discussed above) and their project-related data to conduct a pilot study that investigates the difference between public development project samples and the remaining samples.

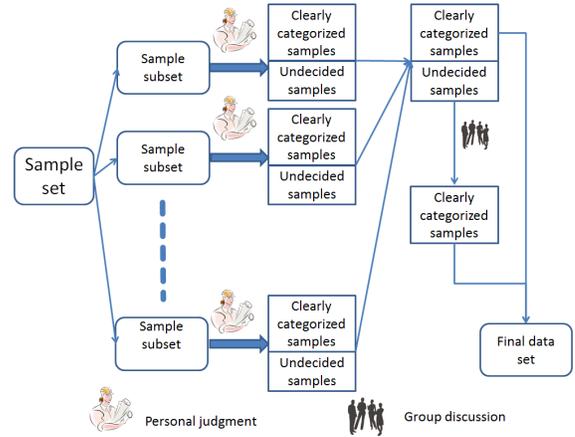

Figure 1. Procedure of classifying projects.

The main findings of this pilot study are twofold (1) most projects (80% in our samples) clearly stated their purposes in the project description. (2) We cannot simply exclude a project that contains some special strings. For example, a project with the word 'test' in its description does not necessarily mean that this project is a private project, and this project may be a 'testing tool'.

The inspiration from this pilot study is that the decision tree model is suitable to address our RQs, because a combination of keywords can avoid the problem in finding (2). For example, if a description contains the keyword 'test', the project may be a private project that aims to test how to use GitHub. But if the description also contains the keyword 'tool', then this project may be a testing tool, and should be classified into public development projects.

In order to improve the keywords that are used in the decision tree, we used the following procedure to gradually improve the quality of matched strings (see Fig. 2):

(1) Generate an initial matching string from the survey samples (1,000 samples discussed above) in the pilot study. For example, we can match projects from other ecosystem by matching string '%mirror%'.
(2) Test the results through the J48 decision tree algorithm on the standard dataset.
(3) Count the number of misclassified samples; if the number exceeds 15% of the standard dataset, investigate these samples and update our strings, and return to (2), else, output the final strings.

As shown in TABLE I, there are several methods to select research samples. Removing projects that are forks is necessary for all researchers who want to study base projects. This method should be used by all studies that aim to investigate base projects, and consequently it is not necessary to compare our model with these methods. Some task-related strategies were used for special research purposes and these methods are not in the scope of our comparison. For example, selecting only

Java projects means that the purpose is to study the pattern that exists merely in Java projects. As our objective is to study a common model to select samples, we should not compare our model with those task-related methods. Hence, our model should be compared with the two methods: (1) removing poor projects and (2) selecting top projects.

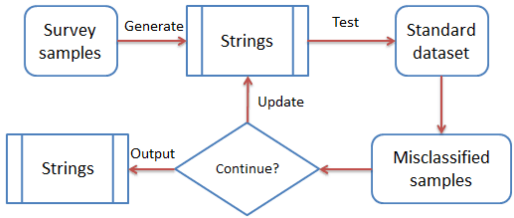

Figure 2 The process of updating string.

There are different dimensions to measure success. We select four dimensions that are widely accepted and easy to obtain [26]: committer numbers, community member numbers, star numbers, and watcher numbers. Then we used different thresholds to segment the dataset to select samples. Methods are shown in Fig. 3.

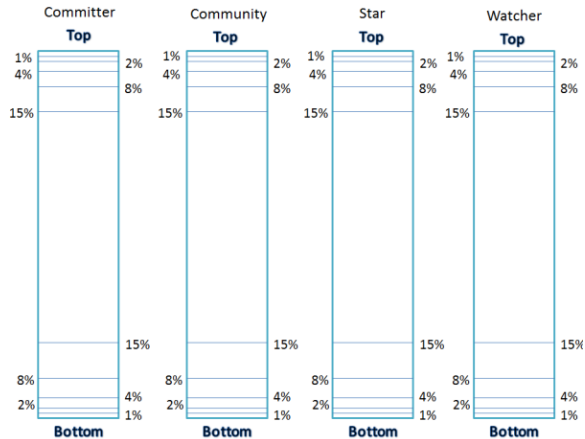

Figure 3 Baseline method schematic diagram.

For each dimension, we have ten choices to select samples (i.e., selecting top 1%,2%,4%,8%,15% projects or deleting bottom 1%,2%,4%,8%,15% projects). Therefore, we should compare 40 different strategies on selecting samples.

## IV. STUDY RESULTS

### A. Standard Dataset

Following the procedure in Section III, we first collected all projects established in GitHub between 2012-1-1 and 2012-1-15. We chose this period of time because GHTorrent started data collection in 2012, and we need to collect about 10,000 projects (considering about the work load). Then, we deleted projects that are forks (recommended by [17, 18]) and obtained 8,638 samples. In addition, we deleted projects that are not described in English or have been removed. Some projects in the GHTorrent dataset have been removed from GitHub by their owners, therefore, we deleted such projects due to the lack of sufficient development information to classify such projects.

Finally, we got 6,715 projects acting as the standard dataset in this study.

Then, we labeled these samples by human judgment. These samples were sent to four participants and we asked them to determine whether these projects are public development projects using our definitions in TABLE II. Doubtful projects (625 projects) were discussed and classified through a meeting. Finally, we obtained 6,715 labeled projects.

### B. Sample Features

After several rounds of iterative process described in Fig. 2. For each sample, we collected keywords existed in the description and URL as features of the sample. Besides, we also collected basic information of projects to help classify samples. All features are shown in TABLE III.

TABLE III. SAMPLE FEATURES

| Feature Source | Feature Content | Remarks |
|---|---|---|
| Description | mirror, fork, moved, longer, test, personal, website, framework, tool, module, component, app, system, dotfiles, collection, blog, plugin, library, server, config, guide, set, repository, deprecated, file, demo, my, github, dot, simple, extension, helper, template, http, https, source, setting, list of, collection of, example, vim, sample, university, school, practice, backup, intro, first, tutorial, course, copy, null, localization, storage, theme, resume, clone, translation, documentation | If the description of a project contains a feature string, this feature is set to 1. If not, this feature is set to 0. For example, project $i$'s desscription contains string 'mirror', then we set feature "mirror" of project $i$ to 1. |
| Basic information | star number, watcher number, community member number, committer number. have language | This information is available through the GitHub API. |
| URL | Dot, config, doc | If URL of a project contain this string, this feature is set to 1. If not, this feature is set to 0. |

### C. Models and Effects

We used the J48 algorithm in Weka[1] tool to fit the model, and debug parameter *confidencefactor*[2] to control model size and results. Then, we got two typical models: a simple model to automatically classify projects and a complex model (in need of human judgment) that can meet strict requirements of studies.

1) Simple model

When *confidencefactor* is set to 0.05, we can get a simple model. The decision tree model is shown in Fig. 4. This model can achieve precision of 0.827 and recall of 0.947 in classifying public development projects. Furthermore, this model is stable, in the sense that precision and recall do not change drastically (precision of 0.820 and recall of 0.941 relatively) when using the 10-fold cross-validation.

2) Complex model

---
[1] https://www.cs.waikato.ac.nz/ml/weka/
[2] A parameter that affects the pruning process of a decision tree. The smaller the parameter value, the smaller the model.

```
simple = 0
|   Tutorial = 0
|   |   dot = 0
|   |   |   have_language = 0: FALSE
|   |   |   have_language = 1
|   |   |   |   mirror = 0
|   |   |   |   |   my = 0
|   |   |   |   |   |   collectionof = 0
|   |   |   |   |   |   |   fork = 0
|   |   |   |   |   |   |   |   personal = 0
|   |   |   |   |   |   |   |   |   url_dot = 0
|   |   |   |   |   |   |   |   |   |   demo = 0
|   |   |   |   |   |   |   |   |   |   |   example = 0
|   |   |   |   |   |   |   |   |   |   |   |   test = 0
|   |   |   |   |   |   |   |   |   |   |   |   |   url_config = 0
|   |   |   |   |   |   |   |   |   |   |   |   |   |   config = 0
|   |   |   |   |   |   |   |   |   |   |   |   |   |   |   blog = 0: TRUE
|   |   |   |   |   |   |   |   |   |   |   |   |   |   |   blog = 1
|   |   |   |   |   |   |   |   |   |   |   |   |   |   |   |   comunity <= 26
|   |   |   |   |   |   |   |   |   |   |   |   |   |   |   |   |   star <= 4
|   |   |   |   |   |   |   |   |   |   |   |   |   |   |   |   |   |   committer <= 1: TRUE
|   |   |   |   |   |   |   |   |   |   |   |   |   |   |   |   |   |   committer > 1: FALSE
|   |   |   |   |   |   |   |   |   |   |   |   |   |   |   |   star > 4: FALSE
|   |   |   |   |   |   |   |   |   |   |   |   |   |   |   comunity > 26: TRUE
|   |   |   |   |   |   |   |   |   |   |   |   |   |   config = 1
|   |   |   |   |   |   |   |   |   |   |   |   |   |   |   watcher <= 5: FALSE
|   |   |   |   |   |   |   |   |   |   |   |   |   |   |   watcher > 5: TRUE
|   |   |   |   |   |   |   |   |   |   |   |   |   url_config = 1
|   |   |   |   |   |   |   |   |   |   |   |   |   |   vim = 0
|   |   |   |   |   |   |   |   |   |   |   |   |   |   |   committer <= 2: FALSE
|   |   |   |   |   |   |   |   |   |   |   |   |   |   |   committer > 2: TRUE
|   |   |   |   |   |   |   |   |   |   |   |   |   |   vim = 1: TRUE
|   |   |   |   |   |   |   |   |   |   |   |   test = 1
|   |   |   |   |   |   |   |   |   |   |   |   |   watcher <= 1
|   |   |   |   |   |   |   |   |   |   |   |   |   |   framework = 0: FALSE
|   |   |   |   |   |   |   |   |   |   |   |   |   |   framework = 1: TRUE
|   |   |   |   |   |   |   |   |   |   |   |   |   watcher > 1: TRUE
|   |   |   |   |   |   |   |   |   |   |   example = 1
|   |   |   |   |   |   |   |   |   |   |   |   watcher <= 13
|   |   |   |   |   |   |   |   |   |   |   |   |   framework = 0: FALSE
|   |   |   |   |   |   |   |   |   |   |   |   |   framework = 1: TRUE
|   |   |   |   |   |   |   |   |   |   |   |   watcher > 13: TRUE
|   |   |   |   |   |   |   |   |   |   demo = 1
|   |   |   |   |   |   |   |   |   |   |   example = 0: FALSE
|   |   |   |   |   |   |   |   |   |   |   example = 1: TRUE
|   |   |   |   |   |   |   |   |   url_dot = 1
|   |   |   |   |   |   |   |   |   |   is_null = 0
|   |   |   |   |   |   |   |   |   |   |   set = 0
|   |   |   |   |   |   |   |   |   |   |   |   config = 0: TRUE
|   |   |   |   |   |   |   |   |   |   |   |   config = 1: FALSE
|   |   |   |   |   |   |   |   |   |   |   set = 1: FALSE
|   |   |   |   |   |   |   |   |   |   is_null = 1: FALSE
|   |   |   |   |   |   |   |   personal = 1: FALSE
|   |   |   |   |   |   |   fork = 1: FALSE
|   |   |   |   |   |   collectionof = 1: FALSE
|   |   |   |   |   my = 1: FALSE (
|   |   |   |   mirror = 1: FALSE
|   |   |   dot = 1: FALSE
|   Tutorial = 1: FALSE
simple = 1: FALSE
```

Figure 4. Simple decision tree model. Nodes with TRUE or FALSE is leaf nodes and TRUE or FALSE is the corresponding judgment on these nodes.

When *confidencefactor* is set to 0.5, we can get a complex model. The decision tree model cannot be shown in this paper due to the space limit, and this decision tree is provided online[3]. The complex model can achieve a precision of 0.837 and a recall of 0.956. The complex model has a distinct characteristic: a large proportion of misclassified samples are located on two leaf nodes. We showed these paths which are the classification conditions to obtain projects in these leaf nodes in TABLE IV.

TABLE IV. THREE DICISION PATHS THAT CONTAIN THE VAST MAJORITY OF MISCLASSIFIED SAMPLES

| Decision Path (common part) | Decision Path (sub-part) | Correct/Incorrect |
|---|---|---|
| Simple = 0; tutorial = 0; dot = 1; have_language = 1; mirror = 0; my = 0; collectionof = 0; fork = 0; personal = 0; url_dot = 0; demo = 0; example = 0; test = 0; url_config = 0; config = 0; blog = 0; plugin = 0; library = 0; framework = 0; star <= 2; sample = 0; source = 0; set = 0; committer <= 2; app = 0 | Committer <= 1; Star > 0; Classify_Result= TRUE | 1,467/355 |
| | Committer > 1; community <= 2; Classify_Result= TRUE | 538/113 |

If we can manually confirm the samples (2,473 out of 6,715, 36.8%) on these nodes, this model achieves a precision of 0.926 and a recall of 0.959. The result can satisfy most of strict requirements of studies.

### D. Comparison with Baseline Method

We compared our model with two base-line methods: selecting top projects and deleting bottom projects. The results are shown in TABLE V and TABLE VI, respectively.

---
[3] https://github.com/sekematerial/S-E-K-E-supplementary-material

TABLE V. RESULTS OF SELECTING TOP PROJECTS

| | top | precision | recall | | top | precision | recall |
|---|---|---|---|---|---|---|---|
| Committer | 1% | 0.761 | 0.011 | Star | 1% | 0.865 | 0.013 |
| | 2% | 0.768 | 0.233 | | 2% | 0.880 | 0.026 |
| | 4% | 0.791 | 0.048 | | 4% | 0.865 | 0.052 |
| | 8% | 0.783 | 0.095 | | 8% | 0.834 | 0.101 |
| | 15% | 0.771 | 0.176 | | 15% | 0.824 | 0.188 |
| Community | 1% | 0.805 | 0.012 | Watcher | 1% | 0.895 | 0.013 |
| | 2% | 0.835 | 0.025 | | 2% | 0.895 | 0.027 |
| | 4% | 0.835 | 0.050 | | 4% | 0.869 | 0.052 |
| | 8% | 0.811 | 0.098 | | 8% | 0.843 | 0.102 |
| | 15% | 0.772 | 0.176 | | 15% | 0.810 | 0.184 |

TABLE VI. RESULTS OF DELETING BOTTOM PROJECTS

| | bottom | precision | recall | | bottom | precision | recall |
|---|---|---|---|---|---|---|---|
| Committer | 1% | 0.656 | 0.988 | Star | 1% | 0.655 | 0.986 |
| | 2% | 0.655 | 0.976 | | 2% | 0.652 | 0.973 |
| | 4% | 0.652 | 0.952 | | 4% | 0.648 | 0.947 |
| | 8% | 0.646 | 0.904 | | 8% | 0.642 | 0.898 |
| | 15% | 0.637 | 0.824 | | 15% | 0.627 | 0.811 |
| Community | 1% | 0.655 | 0.987 | Watcher | 1% | 0.654 | 0.986 |
| | 2% | 0.653 | 0.974 | | 2% | 0.652 | 0.972 |
| | 4% | 0.649 | 0.949 | | 4% | 0.648 | 0.947 |
| | 8% | 0.644 | 0.901 | | 8% | 0.640 | 0.897 |
| | 15% | 0.635 | 0.821 | | 15% | 0.629 | 0.814 |

We can see from the results that compared with our model (0.827 precision and 0.943 recall), these two baseline methods have obvious weaknesses. Selecting top method has a low recall rate, and the precision of deleting bottom method is not very high. Besides, the two methods do not have room for debugging, which means that there are no direct methods to improve the results of the two methods, while by using 63.2% less human effort than manually confirming all samples, our complex model can get a classification result with precision of 0.926 and recall of 0.959. This is an advantage of our approach.

### E. Answers to Research Questions

We formulated two research questions and perform experiments on the standard dataset with 6,715 labeled samples. Our results can answer these research questions:

**Answer to RQ1**: Yes, we can automatically detect public development projects with only descriptions and basic properties of projects (e.g., committer number), and achieve acceptable accuracy (0.827 precision and 0.943 recall).

**Answer to RQ2**: (1) Our model performs better than existing methods. The precision and recall rates are acceptable to be applied to the experiments on GitHub. (2) For studies that have strict requirements on the dataset, our work can reduce 63.2% human resources in selecting samples.

## V. THREATS TO VALIDITY

In this section, we identified several threats to the validity of the study results.

### A. Construct Validity

In this study, we only measured whether a project is a public development project. Thus, the construct validity is whether projects classified as "TRUE" are real public development projects. As the concept **development project** and **public project** are clearly defined and we asked participants if they had any doubt to classify a sample. Doubtful projects were put into the "undecided" set to be further discussed by a group. Then, in the discussion session, all undecided projects are discussed and decided. Hence, the construct validity is limited.

*B. External Validity*

External validity in this study depends on whether the obtained results can be generalized to the GitHub ecosystem, or further other OSS ecosystems. The information used in this study to classify a public development project only contains description, URL, and basic properties of the project, and such information also exists in other OSS ecosystems. We believe that our model can be applied to other OSS ecosystems.

*C. Reliability*

The GHTorrent dataset used in this study is provided by [10], and it is a public and popular dataset for studing OSS development behaviors. Hence, this study can be repliciated using the dataset. At the same time, to mitigate personal bias, we asked participants to avoid classifying samples that may cause disagreements, and then classified these samples through discussion in the data collection procedure. Hence, we believe that our work is relatively reliable.

## VI. CONCLUSIONS

The study aims to develop a model to automatically detect public development projects. The main points of this study are summarized as follows. First, we can automatically detect public development projects with a precision of 0.827 and a recall of 0.943, which is better than existing sample selection methods. Second, by using 63.2% less human effort than manually confirming all samples, we can get better results with a precision of 0.926 and a recall of 0.959, which can meet strict sample requirements.

ACKNOWLEDGMENT

This work is supported by the National Key Research and Development Program of China (Nos. 2017YFB1400602 and 2016YFB0800401), the National Natural Science Foundation of China (Nos. 61572371, 61702377, and 61773175), the Wuhan Yellow Crane Special Talents Program, the CPSF (No. 2015M582272), the Natural Science Foundation of Hubei Province (No. 2016CFB158), and the Fundamental Research Funds for the Central Universities (No. 2042016kf0033).